\documentstyle[aps,prl,preprint,epsfig]{revtex}
\begin{document}
\draft

\preprint{\tighten\vbox{\hbox{\hfil BIHEP-EP1-99-06}}}
\title{Partial Wave Analysis of $J/\psi \to \gamma (K^+K^-\pi^+\pi^-)$}

\author{
J.~Z.~Bai,$^{1}$    Y.~Ban,$^{6}$      J.~G.~Bian,$^{1}$  A.~D.~Chen,$^{1}$
G.~P.~Chen,$^{1}$   H.~F.~Chen,$^{2}$  H.~S.~Chen,$^{1}$  J.~C.~Chen,$^{1}$
X.~D.~Chen,$^{1}$   Y.~Chen,$^{1}$
% ...
\\(BES Collaboration)\cite{besjpsi}\\
D.~V.~Bugg $^{12}$ and B.~S.~Zou $^{1,12}$}

\vspace{1cm}

\address{
$^{1}$ Institute of High Energy Physics, Beijing 100039,
People's Republic of China \\
$^{2}$ University of Science and Technology of China, Hefei 230026,
People's Republic of China \\
$^{3}$ Shandong University, Jinan 250100,
People's Republic of China \\
$^{4}$ Hangzhou University, Hangzhou 310028,
People's Republic of China \\
$^{5}$ Shanghai Jiaotong University, Shanghai 200030,
People's Republic of China \\
$^{6}$ Peking University, Beijing 100871,
People's Republic of China \\
$^{7}$ Hua Zhong Normal University, Wuhan 430079,
People's Republic of China \\
$^{8}$ China Center for Advanced Science and Technology(CCAST), World
Laboratory, Beijing 100080, People's Republic of China)\\
$^{9}$ Nankai University, Tianjin 300071,
People's Republic of China \\
$^{10}$ Henan Normal University, Xinxiang 453002,
People's Republic of China \\
$^{11}$ Wuhan University, Wuhan 430072,
People's Republic of China \\
$^{12}$ Queen Mary and Westfield College, London E1 4NS, United Kingdom}


\maketitle

\begin{abstract}
BES data on $J/\psi \to \gamma (K^+K^-\pi^+\pi^-)$ are presented.
The $K^*\bar K^*$ contribution peaks strongly near threshold.
It is fitted with a broad $0^{-+}$ resonance with mass
$M = 1800 \pm 100$ MeV, width $\Gamma = 500 \pm 200$ MeV. 
A broad $2^{++}$ resonance peaking at 2020 MeV is also required with width 
$\sim 500$ MeV. There is further evidence for a $2^{-+}$ component peaking
at 2.55 GeV. The non-$K^*\bar K^*$ contribution is close to phase space; 
it peaks at 2.6 GeV and is very different from $K^{*}\bar{K^{*}}$.
\end{abstract}

\vspace{0.6cm}

In 1990, MARK III presented their spin-parity analysis
of $J/\psi \rightarrow \gamma K^{*}\bar{K^{*}}$ at the Rheinfels Workshop
on the Hadron Mass Spectrum \cite{mark3}. They found a dominant $0^{-+}$
component, accounting for $55\%$ of the data. In addition, smaller 
$2^{+}$ and $2^{-}$ channels of about equal strength were observed.
Interferences were not included. Here we present BES data and carry out
a full partial wave analysis.

Lattice QCD and other theoretical models predict $0^{-+}$ and $2^{++}$
glueballs with masses 2.0 to 2.4 GeV \cite{Bali}. Recently, several broad
structures have been identified in this mass range. In an analysis of
Mark III data on radiative decays of $J/\psi$ to $\eta \pi^+\pi^-$,
$\rho\rho$, $\omega\omega$, $K^*\bar K^*$ and $\phi\phi$, a very broad
$0^-$ component has been found \cite{b0m} with a mass of 1750--2190 MeV
and a width of order 1 GeV. Its decays are flavour blind to first
approximation, making it a candidate for the $0^-$ glueball.
This $0^-$ component appears strongly in BES data on 
$J/\psi \rightarrow \gamma (\pi^{+} \pi^{-} \pi^{+} \pi^{-})$ \cite{besg4p}. 
There, we also find a broad $2^{+}$  contribution fitted as a resonance 
at $1940 \pm 60$ MeV with $\Gamma = 350 \pm 100$ MeV. The WA91 and WA102 
groups have
also found a broad $2^{+}$ resonance at 1920 MeV in their $4\pi$ mass 
spectrum\cite{wa102}. A recent partial wave analysis of Crystal Barrel
data on $p\bar{p} \to \eta\eta \pi^0$ \cite{dvb} has found a broad
$f_2(1980)\to\eta\eta$ resonance with mass 1980 MeV and width 500 MeV.
There are further data on $\bar pp \to \eta\pi^0 \pi^0$ \cite{cb-3b,epp}, 
where a broad $0^-$ component is observed plus a peak in $\eta\sigma$
at 2140 MeV with width $\sim 150$ MeV. In Refs. \cite{b0m} and \cite{besg4p},
an $f_0(2100)$ was observed in $4\pi$ states. A peak at this mass was observed
earlier in the E760 experiment \cite{E760}. The $f_0(2100)$ decay to
$\eta\eta$ has been confirmed in two sets of Crystal Barrel 
data \cite{eep,ee}.

The analysis in this paper uses $7.8 \times 10^6 ~J/\psi$ triggers
collected by the Beijing Spectrometer(BES). This detector has been described
in detail in Ref \cite{detect}. Here we describe briefly those detector
elements crucial to this measurement. Tracking is provided by a 10 superlayer
main drift chamber (MDC). Each superlayer contains four layers of sense wires
measuring both the position and the ionization energy loss ($d$E$/dx$) of
charged particles. The momentum resolution is
$\sigma_P/P = 1.7\%\sqrt{1 + P^2}$, where $P$ is the momentum of charged
tracks in GeV/$c$. The resolution of the $d$E$/dx$ measurement is about
$9\%$. This provides good $\pi/K$ separation and proton identification in the
low momentum region. An array of 48 scintillation counters surrounding the
MDC measures the time-of-flight (TOF) of charged tracks with a resolution of
330$ps$ for hadrons. Outside the TOF system is an electromagnetic calorimeter
composed of streamer tubes and lead sheets with a $z$ positional resolution
of 4 cm. The energy resolution scales as $\sigma_E/E = 22\%/\sqrt{E}$, where
$E$ is the energy in GeV. Outside the shower counter is a solenoidal magnet
producing a 0.4 Tesla magnetic field.

Each candidate event is required to
have exactly four charged tracks. Every track must have a good helix fit
in the polar angle range $-0.8 < \cos\theta < 0.8$ and a transverse
momentum $>60$ MeV/c.
A vertex is required within an interaction region $\pm 20$ cm longitudinally
and 2 cm radially. At least one reconstructed $\gamma$ is required in the
barrel shower counter. A minimum energy cut of 80 MeV is imposed on the
photons. Showers associated with charged tracks are also removed.

A positive identification of at least one $K^{\pm}$ and one $\pi ^{\pm}$ is
required using time of flight and/or $dE/dx$. If two tracks are ambiguous,
both alternative identifications are tried. Events are fitted kinematically
to the 4C hypothesis $J/\psi \to \gamma (K^+K^-\pi^+\pi^-)$, requiring a
confidence level $>5\%$. If there is more than one photon, the fit is
repeated using all permutations.  Events with two or more photons are also
fitted to $J/\psi \to \gamma\gamma K^+K^-\pi^+\pi^-$. Those giving a better
fit than to $\gamma (K^+K^-\pi ^+\pi ^-)$ are rejected, as are events
fitting the final state $K^+K^-\pi ^+\pi ^-$. Next, we require
$\mid{U_{miss}}\mid = \mid E_{miss}-P_{miss} \mid <0.12$ GeV/c$^2$,
so as to reject the events with multi-photons or more or less than two
charged kaons; here, $E_{miss}$ and $P_{miss}$ are, respectively, the
missing energy and missing momentum of all charged particles; they are
calculated by assuming the charged particles are $K^{+}K^{-}\pi^{+}\pi^{-}$.
The momentum of the $K^{+}K^{-}\pi^{+}\pi^{-}$ system transverse to the photon 
$P_{t\gamma}^2=4\mid{P_{miss}}\mid{^2}~\sin^2(\theta_{m\gamma}/2) <0.006$
(GeV/c)$^2$ is required in order to remove the background $J/\psi\rightarrow
\pi^0K^{+}K^{-}\pi^{+}\pi^{-}$; here $\theta_{m\gamma}$ is the angle between
the missing momentum and the photon direction. Finally, to remove a small
background of $K_s^0$, we perform a cut on the $\pi^+\pi^-$ invariant mass,
$\mid M_{\pi^+\pi^-}-M_{K_s^0} \mid > 25$ MeV; to remove the small background
due to $\phi(1020)$, we use a cut on the $K^+K^-$ invariant mass,
$\mid M_{K^+K^-}-M_{\phi} \mid > 20$ MeV;
background $J/\psi\rightarrow\omega K^+ K^-$ events are eliminated by the cut
$\mid M_{\pi^+\pi^-\pi^0}-M_{\omega} \mid > 25$ MeV, in fitting the
$\pi^{0} K^{+} K^{-}\pi^{+}\pi^{-}$ hypothesis with only one photon
detected and the $\pi^{0}$ associated to the missing momentum.
The number of surviving events is 1516.

The effects of the various selection cuts on the data is simulated with a full
Monte Carlo of the BES detector; 500,000 Monte Carlo events are successfully
fitted to $J/\psi \to \gamma (K^+K^-\pi^+\pi^-)$; all background reactions are
similarly fitted to this channel. The estimated background is 16\%, purely from 
$J/\psi \to \pi ^0 (K^+K^-\pi ^+\pi ^-)$. We have included this background in
the amplitude analysis, but it lies very close to non-$K^*\bar K^*$ events and
has negligible effect on the analysis of $K^*\bar K^*$ events. 
                    
Fig.~\ref{figure1} shows the $KK$, $\pi\pi$, $K\pi$, $KK\pi$, $K^* K$ and
$K^* \pi$ invariant mass distributions of selected events after
acceptance cuts. 
These cuts are accomodated in the maximum likelihood fit, using the full
Monte Carlo simulation of the detector. 
A strong $K^*$ signal is clearly visible. 
There is just a hint of possible structure in Fig.~\ref{figure1}(a) around
1500 MeV, but including $f_0(1500)$ or $f_2'(1525)$ into the fit fails to
reveal any significant contribution. It is too narrow for $f_2'(1525)$.
Figs.~\ref{figure1}(e) and (f) show no
significant structure attributable to heavier $K_1$ or $K^*$'s states.
The $K^{+}K^{-}\pi^{+}\pi^{-}$ invariant mass distribution of 
$J/\psi\rightarrow\gamma K^{+}K^{-}\pi^{+}\pi^{-}$ is shown in
Fig.~\ref{figure2}(a). The $K^{*}\bar{K^{*}}$ mass spectrum is obtained with
a double cut on the two $K\pi$ invariant masses,
$\mid M_{K\pi}-M_{K^{*}} \mid \,< 75$ MeV and is shown in
Fig.~\ref{figure2}(b). Fig.~\ref{figure3}(a) shows the
$K^{+} K^{-}\pi^{+}\pi^{-}$ mass spectrum of
$J/\psi\rightarrow \gamma K^{+}K^{-}\pi^{+}\pi^{-}$ when one
$K\pi$ or both $K\pi$ masses fall outside the $K^{*}$ region.

We first attempted to fit the non-$K^*\bar K^*$ data of
Fig.~\ref{figure3}(a) as $\kappa\kappa$ and $K^{*0}\kappa + c.c.$;
here $\kappa$ means the $K\pi$ S-wave. The $\kappa$ is parametrized to fit
LASS data \cite{LASS} on $K^{*}_{0}(1430)$. Each $\kappa$ is made from
$K^+\pi ^-$ or $K^-\pi ^+$, and if the $\kappa$ really dominates this process,
the $K\pi$ mass distribution should follow it. The dashed curve in
Fig.~\ref{figure3}(b) shows that this fails.

A simple solution which works well is to fit instead with $K^+K^-\pi^+\pi^-$
phase space. This is shown by the full curve of Fig.~\ref{figure3}(b).
A detailed study of non-$K^*\bar K^*$ events reveals only a weak $\rho (770)$
signal, visible in Fig.~\ref{figure1}(b). However, when we perform a cut on
the $\pi ^+\pi ^-$ mass: $|M_{\pi \pi} - M_{\rho } |< 100$ MeV, we have been
unable to correlate this signal with any particular
$K^+K^-$, $K^{\pm}\pi ^+\pi ^-$ or $K^+K^-\pi ^+\pi ^-$ resonance.

We now turn to $K^*\bar K^*$ events. One sees immediately a large difference
between $K^*\bar K^*$ and non-$K^*\bar K^*$ events from
Fig.~\ref{figure2}(b) and \ref{figure3}(a). The $K^*\bar K^*$ events peak
strongly close to threshold. We shall show that the data require a strong
$0^-$ peak at threshold. 

The amplitudes in the PWA analysis are constructed from Lorentz-invariant
combinations of the 4-vectors and the photon polarization for $J/\psi$
initial states with helicity $\pm 1$. Cross sections are summed over photon
polarizations. The relative magnitudes and phases of the amplitudes are
determined by a maximum likelihood fit. We use $\ell $ to denote the orbital
angular momentum between the photon and $KK\pi\pi$ states in the
production reaction. Because this is an electromagnetic transition, the same
phase is used for amplitudes with different $\ell $ but otherwise the same
final state. In fitting $K^*\bar K^*$, spin-parity assignments up to $J = 4$
have been tried. 

We have examined slices of the $K^*\bar{K^*}$ mass spectrum $\sim 200$ MeV
wide. Each slice has been fitted with a constant contribution with quantum
numbers $0^{++}$, $0^{-+}$, $1^{++}$, $2^{++}$, $2^{-+}$ and $4^{++}$. 
These different quantum numbers give angular distributions which are
distinctively different. They depend on two angles (a) $\theta _{K^+}$ or
$\theta _{\pi^+}$ of $K^+$ or $\pi^{+}$ with respect to the $K^{+}\pi^{-}$ or
$K^{-}\pi^{+}$ pair in their rest frame, and (b) the azimuthal angle $\chi$
between the planes of $K^{+}\pi^{-}$ and $K^{-}\pi^{+}$ in the rest frame
of the resonance. Results of the slice fit are shown in Fig.~\ref{figure4}.
The $0^{-+}$ contribution is largest and peaks towards the low mass end.
There is evidence for a broad $2^{++}$ contribution peaking at
$\sim 2050$ MeV, and this will be confirmed below by detailed fits. 
Contributions from $0^{++}$ and $4^{++}$ are small or absent.
Though the results of the slice fit show that there may be some $1^{++}$
contributions in the  $2.0 \sim 2.4$ GeV mass region and some $2^{-+}$ 
contributions at low mass, there is some degree of cross-talk with 
$0^-$ and $2^+$ signals; when we put  $1^{++}$ or $2^{-+}$ into the full fit
in these regions, their contributions are found to be negligible and are
absorbed into $0^-$ or $2^+$. At high masses, there is some evidence for a
$2^{-+}$ contribution.

The precise mass dependence of each contribution is hard to establish
because signals are broad and are affected by systematic uncertainties
as follows.
(a) There is an unknown form factor for the vertex $J/\psi \to \gamma X$.
For consistency with other work, we adopt a very weak form factor 
$\exp (-\alpha _gq^2)$ with
$\alpha _g = 0.13 $ GeV$^{-2}$, fitted to many channels of $J/\psi$ radiative
decay \cite{broad}; $q$ is the photon momentum in the $J/\psi$ rest frame.
(b) Likewise we use a form factor $\exp (-\alpha_d p^2)$ with $\alpha_d = 2.0$
GeV$^{-2}$, where $p$ is the momentum of the final $K^*$ in the rest frame
of $X$;
(c) The resonance $X$ may couple to further channels at higher mass; this
would pull the cross section down when these channels open.
The effect of these form factors is to allow shifts of $\pm 50$ MeV  in the
mass fitted to the broad component; however, fitted intensities hardly
change in shape at all. 

We begin by fitting simple Breit-Wigner resonances of constant width to
channels $0^{-+}$, $2^{++}$ and $2^{-+}$, including the form factors 
discussed above. Appropriate Blatt-Weisskopf centrifugal barrier factors are
included with a radius consistent with the $\alpha$ parameters. The best fit
is shown on Fig.~\ref{figure5}. Crosses are data and histograms the fit.

The optimum mass for the broad $0^-$ contribution is $1800 \pm 100$ MeV.
The width optimises at $500 \pm 200$ MeV. 
The present fit with a simple Breit-Wigner form is likely to be an 
over-simplification for such a broad state. In Ref. \cite{broad}, all
radiative decays are fitted simultaneously. This produces a slightly
broader peak, shown by the dashed curve in Fig.~\ref{figure5}(b), but with 
little effect on other components.

The $2^+$ contribution of Fig.~\ref{figure5}(c) is definitely required.
Without it, log likelihood gets worse by 58; this is roughly an $8\sigma$
effect, from tables of log likelihood for 6 fitted parameters. 
The $2^-$ contribution is less secure. It improves log likelihood
by 18.9 ($\sim 4.7 \sigma )$. Both the $2^-$ and $0^-$ contributions are 
produced in a P-wave and consequently vary as $q^3$, where $q$ is the momentum
of the photon;
this dependence leads to a slight underfit at the highest masses.
If this defect is real, it is most likely fitted by a small E1 cross section
for production of high mass $0^+$ or $2^+$ components, which cannot be
identified definitively.

The broad $2^+$ component near 2 GeV is interesting. It is very close to
that required to fit BES data on $J/\psi \to \gamma (\pi^+\pi^-\pi^+\pi^-$).
It peaks at 2020 MeV. Because the $K^*\bar{K^*}$ phase space opens rapidly, 
the pole position tends to come out lower, around 1900 MeV with a width 
of 500 MeV. Changing the form factors moves the pole position up or down
slightly, but leaves the fitted intensity almost unchanged.

Figs.~\ref{figure6}(a) and (b) shows fitted angular distributions, 
summed over all $K^*\bar K^*$ masses. Figs.~\ref{figure6}(c) and (d) 
illustrate angular distributions from individual quantum numbers. 
The necessity for a large $0^-$ contribution is obvious. 
The full angular dependence involves correlations between
$\theta$ and $\chi$ (defined earlier), containing much more information 
than the projections of Fig.~\ref{figure6}.
                                                                          
No significant $0^{++}$, $1^{++}$ or $4^{++}$ resonance is found in the full
fit (changes in log liklihood $< 2.5$).
That result is itself  interesting. There is no evidence for the $f_0(2100)$,
which appears strongly in $J/\psi \to \gamma (4\pi)$ and also in two sets of
Crystal Barrel data in the $\eta \eta$ channel and also in E760 data.
Its absence from the present data suggests it is not an $s\bar s$ state.

Table~\ref{table1} summarises branching ratios evaluated from the fit; 
the first error is statistical
and the second covers systematic errors in the overall normalisation of the
number of $J/\psi$ interactions. The $\eta_c$ component will be the subject of
a separate publication, and has been subtracted. 
We find 1516 $KK\pi\pi$ events (of which 320 are
$K^*\bar K^*$) within the cuts we have applied. For each $K^*$, 17.5\% of
events lie outside cuts. We correct branching ratios to allow for this loss. 
We also correct for all charges in $K^*\bar K^*$ channels:
we have multipied values obtained here by a factor $(1.5^2)$ for
Clebsch-Gordan coefficients in $K^*$ decay and a further factor
2 for production of $K^{*+}\bar K^{*-}$ in the radiative decay process.

We have also searched for a $2^-$ contribution. 
A component with these quantum numbers was found in
$J/\psi \to \gamma (\eta \pi^+ \pi^-)$ at 1840 MeV \cite{eta2pi}.
A possible interpretation is as an $I = 0$ hybrid; an $s\bar sg$ hybrid is
then to be expected around 2.1 GeV with roughly half the branching ratio,
summed over all kaonic channels. 
We find that such a $2^-$ component with width 250--400 MeV produces only
an insignificant improvement in the fit: an improvement in log likelihood
of 4.5 for 4 extra parameters. 
The branching ratio for production and decay of the fitted $2^-$ component, 
corrected for all charge states, is then $1.8 \times 10^{-4}$. 
This is to be compared with the value $(9.3 \pm 3.33)\times 10^{-4}$ observed
in all $\eta \pi \pi$ channels.  

We summarize as follows.
We fit with a broad $0^{-+}$ resonance with $M = 1800 \pm 100$ MeV and 
$\Gamma = 500 \pm 200$ MeV, decaying to $K^{*}\bar{K^{*}}$.
It is to be identified with the broad $0^-$ component found also
in $\eta \pi^+\pi^-$, $\rho\rho$, $\omega\omega$ and $\phi\phi$.
The data also definitely required a broad $2^{++}$ resonance $f_2(1950)$
decaying to $K^{*}\bar{K^{*}}$.
The fit to the high mass end of the $K^+K^-\pi ^+\pi ^-$ spectrum
is best achieved with an additional  $2^{-+}$ signal. 

The BES group thanks the staff of IHEP for technical support in running the
experiment. This work is supported in part by China Postdoctoral Science
Foundation and National Natural Science Foundation of China under contract
Nos. 19991480, 19825116 and 19605007;
and by the Chinese Academy of Sciences under contract No. KJ 95T-03(IHEP).
We also acknowledge financial support from the Royal Society
for collaboration between Chinese and UK groups.

\begin{center}

\begin{figure}[htbp]
\centerline{\epsfig{file=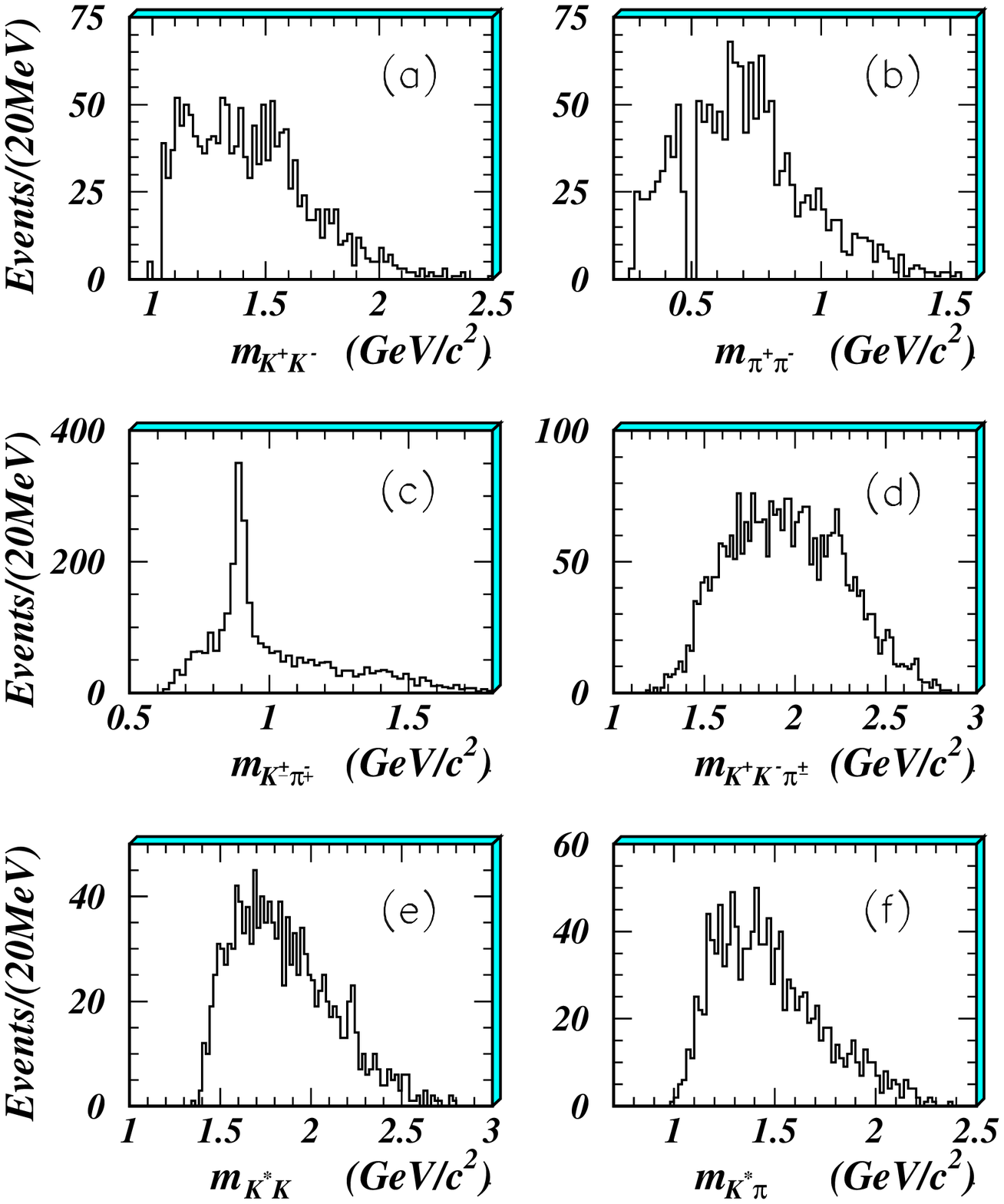,width=5.0in,height=5.8in}} 
\caption[]{The $KK$, $\pi\pi$, $K\pi$ (two entries/event),
$KK\pi$ (two entries/event), $K^*K$ and $K^*\pi$  masses of
$J/\psi\rightarrow\gamma(K^{+}K^{-}\pi^{+}\pi^{-})$.}
\label{figure1}
\end{figure}

\newpage
\begin{figure}[htbp]
\centerline{\epsfig{file=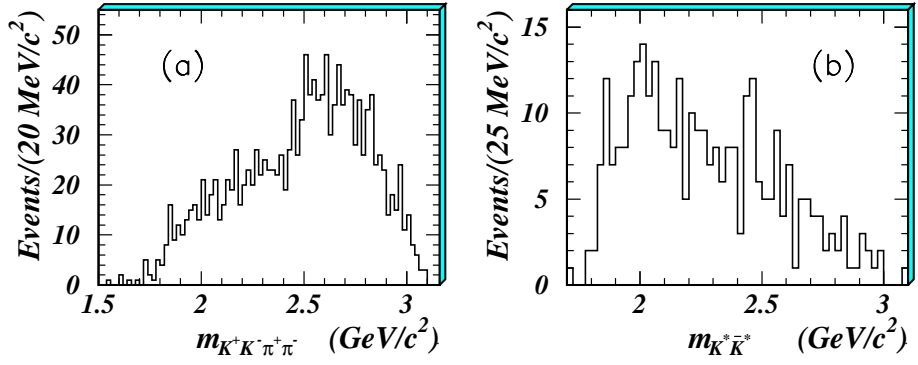,width=5.0in,height=2.4in}} 
\caption[]{(a) The $KK\pi\pi$ mass of 
$J/\psi\rightarrow\gamma K^{+}K^{-}\pi^{+}\pi^{-}$; 
(b) The $K^{*}\bar{K^{*}}$ mass of 
$J/\psi \rightarrow \gamma K^{*}\bar{K^{*}}$}
\label{figure2}
\end{figure}

\begin{figure}[htbp]
\centerline{\epsfig{file=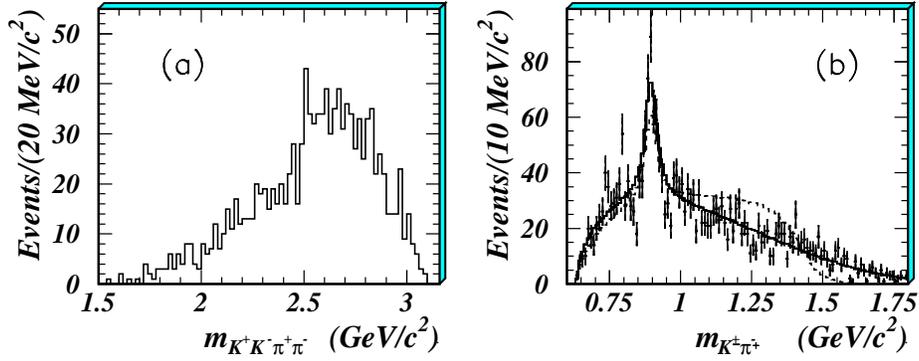,width=5.0in,height=2.4in}} 
\caption[]{ The (a) $KK\pi\pi$ and (b) $K\pi$ mass 
spectrum from non-$K^*\bar K^*$ events. In (b), the dashed line
shows the $\kappa$ fitted to $K^*_0(1430)$ and the full line is
derived from $K^+K^-\pi ^+\pi ^-$ phase space plus a fit to
$K^* + K\pi$ phase space.}
\label{figure3}
\end{figure}

\begin{figure}[htbp]
\centerline{\epsfig{file=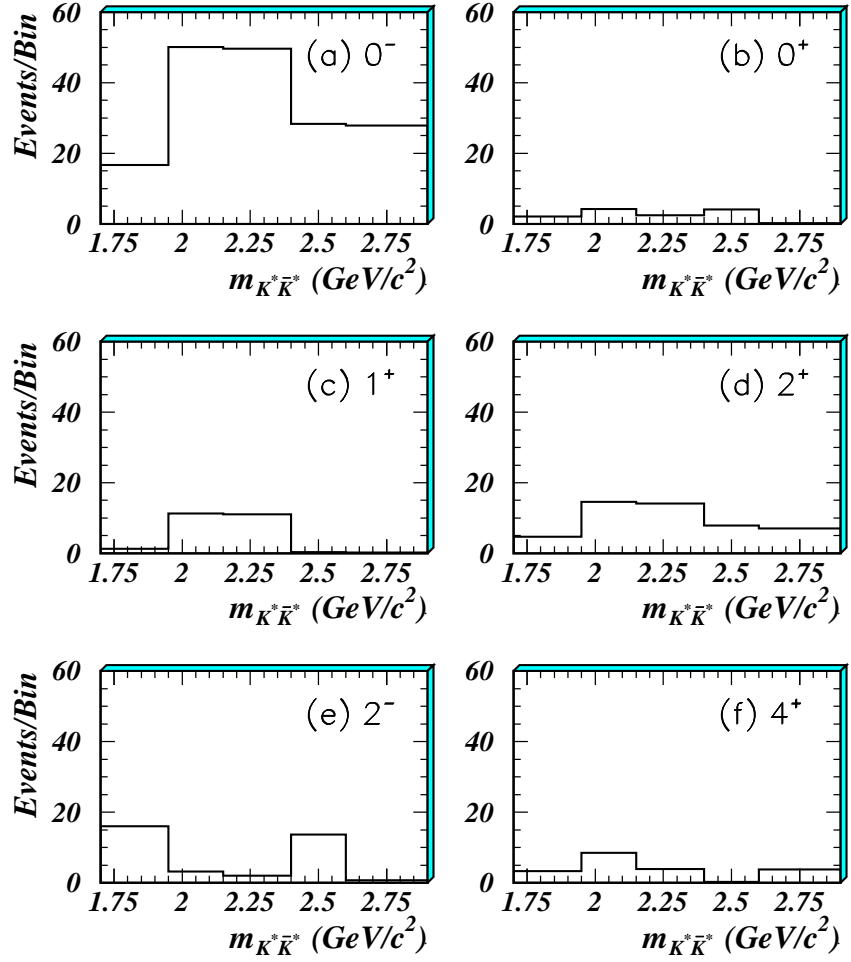,width=4.5in,height=5.3in}}
\caption[]{The contribution of (a) $0^-$, (b) $0^+$, (c) $1^+$,
(d) $2^+$, (e) $2^-$ and (f) $4^+$ from the slice fit to 
$J/\psi \rightarrow \gamma K^{*}\bar{K^{*}}$}
\label{figure4}
\end{figure}

\begin{figure}[htbp]
\centerline{\epsfig{file=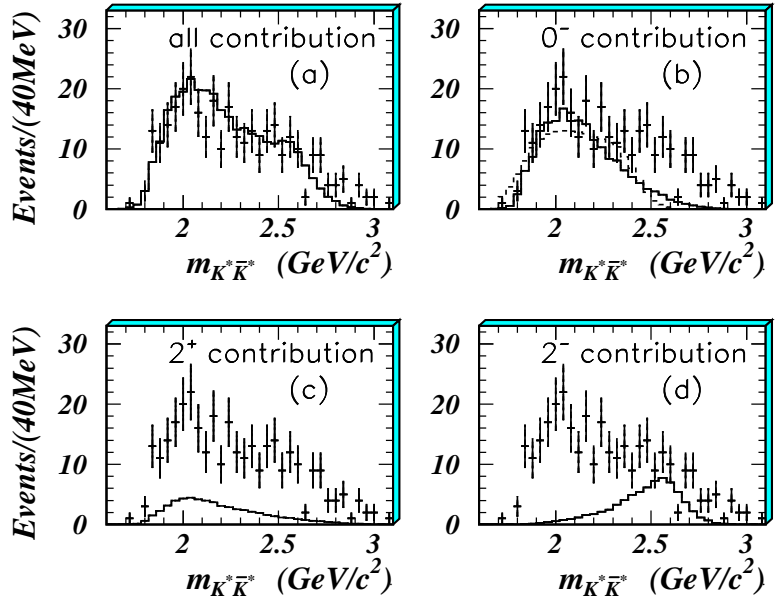,width=4.5in,height=3.55in}} 
\caption[]{Comparison of the mass spectrum of $K^*\bar{K^*}$ with
(a) all, (b) $0^-$, (c) $2^+$ and (d) $2^-$ contributions
from the best fit; the dashed curve shows the $0^-$ component
fitted to the broad $0^-$ resonance in Ref. \cite{broad}.}
\label{figure5}
\end{figure}

\begin{figure}[htbp]
\centerline{\epsfig{file=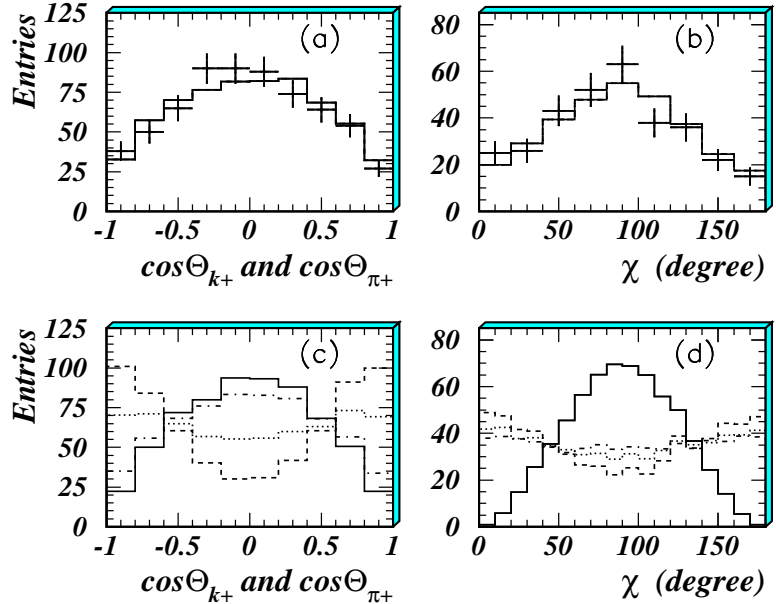,width=4.5in,height=3.55in}} 
\caption[]{(a) Comparison of $J/\psi\rightarrow\gamma K^*\bar{K^*}$ 
data and fit for (a) $\cos \theta _{K^+} + 
\cos \theta _{\pi^+}$ (both entries summed) and (b) $\chi$; 
(c) and (d) corresponding
angular distributions for $0^-$ (full histograms), $0^+$ (dashed), 
$2^+$ (dotted) and $2^-$ (dash-dot).}
\label{figure6}
\end{figure}   

\begin{table}[htbp]
\begin{minipage}[]{0.9\linewidth}
\caption{Branching ratios (BR) integrated over all masses, but
excluding the $\eta_c$.}
\label{table1}
\vspace{0.35cm}
\begin{tabular}{|cc|} 
\hline
Channel & BR \\
\hline
$BR(J/\psi \to \gamma K^+K^-\pi ^+\pi ^-)$  
& $(2.1 \pm 0.1 \pm 0.6) \times 10^{-3}$ \\
$BR(J/\psi \to \gamma K^{*}\bar K^{*})$  
& $(4.0 \pm 0.3 \pm 1.3) \times 10^{-3}$ \\
$BR(J/\psi \to \gamma 0^-) \times BR(0^- \to K^{*}\bar K^{*})$  
& $(2.3 \pm 0.2 \pm 0.7) \times 10^{-3}$ \\
$BR(J/\psi \to \gamma f_2(1950)) \times BR(f_2(1950) \to  K^{*}\bar K^{*})$  
& $(0.7 \pm 0.1 \pm 0.2) \times 10^{-3}$ \\
$BR(J/\psi \to \gamma 2^-) \times BR(2^- \to  K^{*}\bar K^{*})$  
& $(0.9 \pm 0.1 \pm 0.3) \times 10^{-3}$ \\
\hline
\end {tabular}
\end{minipage}
\end{table}

\end{center}

\end{document}